%% file: NuPhys_okeeffe.tex
\documentclass[12pt]{article}
\usepackage{graphicx}

\newcommand\pubnumber{SNSN-323-63}
\newcommand\pubdate{\today}

\def\napoli{Physics Department\\
Lancaster University, Lancaster, LA1 4YB, UK}
\def\support{\footnote{On behalf of the T2K collaboration.}}

\def\Title#1{\begin{center} {\Large #1 } \end{center}}
\def\Author#1{\begin{center}{ \sc #1} \end{center}}
\def\Address#1{\begin{center}{ \it #1} \end{center}}

\newcommand\pubblock{\rightline{\begin{tabular}{l} \pubnumber\\
         \pubdate  \end{tabular}}}
\newenvironment{Abstract}{\begin{quotation}  }{\end{quotation}}
\newenvironment{Presented}{\begin{quotation} \begin{center} 
             PRESENTED AT\end{center}\bigskip 
      \begin{center}\begin{large}}{\end{large}\end{center} \end{quotation}}

\input econfmacros.tex
\begin{document}
\begin{titlepage}
\pubblock

\vfill
\Title{Current status and near future plans for T2K}
\vfill
\Author{ Helen Mary O'Keeffe\support}
\Address{\napoli}
\vfill
\begin{Abstract}

\end{Abstract}
\vfill
\begin{Presented}
NuPhys 2015: Prospects in Neutrino Physics\\
London, UK, December 16-18, 2015.
\end{Presented}
\vfill
\end{titlepage}
\def\thefootnote{\fnsymbol{footnote}}
\setcounter{footnote}{0}

\section{Introduction}
The Tokai to Kamioka (T2K) experiment is a long-baseline neutrino oscillation experiment located in Japan \cite{T2K}.  An intense, almost pure beam of $\nu_{\mu}$ is produced at the J-PARC facility in Tokai-mura,  by colliding 30 GeV protons with a stationary graphite target.  This produces a beam of secondary hadrons, mainly pions and kaons, from which pions are selected using a series of three magnetic horns.  The selected pions decay in a 96 m long decay volume.  Depending on the polarity of the horn current, $\pi^+$ or $\pi^-$ can be chosen to produce a $\nu_{\mu}$ or $\bar{\nu}_{\mu}$ beam respectively. The neutrino beam is directed $2.5^{\circ}$ away from the axis between the target and the far detector 295\,km away.  This off-axis technique produces a narrow band beam with a peak energy around 0.6\,GeV.  This corresponds to the energy of the first $\nu_{\mu} \rightarrow \nu_e$ oscillation maximum for a baseline of 295 km. 

The near detector complex, located approximately 280 m from the target, consists of an on-axis Interactive Neutrino GRID (INGRID) detector and an off-axis near detector (ND280).  The primary purpose of the INGRID detector is to measure the direction, stability and flux of the on-axis beam.  The ND280 detector is $2.5^{\circ}$ off-axis and comprises of five sub-detectors, namely a $\pi^0$ detector, two active fine grained detectors, three gaseous argon time projection chambers, an electromagnetic calorimeter and a side muon range detector.  These sub-detectors are located inside a magnet, which provides a 0.2 T field for charge identification.  The ND280 detector is used for measurements of interaction cross-sections and the neutrino flux, in particular measurement of the intrinsic electron (anti)neutrino content of the beam.

The far detector is the Super-Kamiokande water Cherenkov detector, located 295 km away from the neutrino production point \cite{SK}.  Super-Kamiokande is divided into an inner and outer detector.  The inner detector has a 22.5\,kton water fiducial volume surrounded by 11,129 photomultiplier tubes (PMTs).  A 2\,m wide outer detector surrounds the inner detector and PMTs.  In the water, neutrinos interact to produce their corresponding charged lepton partner which, if sufficiently energetic, produce Cherenkov light in the water.   Good separation between $\nu_e$ and $\nu_{\mu}$ candidates is achieved via a particle identification variable, with a probability of misidentifying a $\mu$ as an $e$ of $<1$\%. 

T2K was optimised to perform a precision measurement of $\theta_{23}$ and $\Delta m^2_{32}$ via $\nu_{\mu}$ disappearance and to search for the mixing angle $\theta_{13}$ via $\nu_e$ appearance in the far detector.  Having confirmed the appearance of $\nu_e$ in a $\nu_{\mu}$ beam in 2013 \cite{NUE}, T2K has been taking data in anti-neutrino mode.  To June 2015, $\sim 7.0 \times 10^{20}$ POT in neutrino mode and $\sim4.0 \times 10^{20}$ POT in anti-neutrino mode have been taken.  

\section{Recent results from T2K}
In neutrino mode, the T2K experiment has measured $\theta_{13}$ via $\nu_e$ appearance in a $\nu_{\mu}$ beam \cite{NUE} and also $\theta_{23}$ via $\nu_{\mu}$ disappearance \cite{NUMU}.  Approximately 1\% of the initial $\nu_{\mu}$ beam is $\nu_e$, which is the most significant background for the $\nu_e$ appearance measurement.   This intrinsic $\nu_e$ content and other backgrounds e.g. misidentified $\pi^0$ events at Super-Kamiokande, result in an expected background of $4.64 \pm 0.53$ electron-like events in the far detector.  With the full data set of $6.57 \times 10^{20}$ POT, a total of 28 electron-like candidates were observed, a significance of $7.3 \sigma$.  This was the world first measurement of electron neutrino appearance in a muon neutrino beam \cite{NUE}. 

In the $\nu_{\mu}$ disappearance search, a clear deficit in the number of muon-like events was observed at Super-Kamiokande.  For the full neutrino data-set of $6.57 \times 10^{20}$ POT, 120 $\nu_{\mu}$ candidates were observed, significantly fewer than the $446\pm 22.5$ expected without oscillation \cite{NUMU}.   This led to a world-leading measurement of $\theta_{23}$ of $\sin^2\theta_{23} = 0.514^{+0.055}_{-0.056}$ assuming a normal hierarchy and $\sin^2\theta_{23} = 0.511^{+0.055}_{-0.055}$ assuming an inverted hierarchy.  Both values are in good agreement with similar measurements from the MINOS experiment \cite{MINOS} and Super-Kamiokande's atmospheric results \cite{SK}. 

By performing a joint fit to $\nu_e$ appearance and $\nu_{\mu}$ disappearance and combining the T2K result with the world average value of $\theta_{13}$ from reactor experiments, $\delta_{CP}$ between $0.19\pi$ and $0.80\pi$ are excluded at 90\% C.L.~for the normal hierarchy. For the inverted hierarchy values between $-\pi$ and $-0.97\pi$ and $-0.04\pi$ and $\pi$ are excluded at 90\% C.L. \cite{COMB}.

Since 2014, T2K has been running with an $\bar{\nu}_{\mu}$ beam, allowing a search for $\bar{\nu}_{\mu}$ disappearance and also $\bar{\nu}_e$ appearance.  For the $\bar{\nu}_{\mu}$ disappearance search, a clear deficit in the number of muon-like events has been detected at Super-Kamiokande.  For $4.0 \times 10^{20}$ POT, 34 events were observed, significantly fewer than the 103.6 events expected for the no oscillation hypothesis, leading to the world's most precise value of $\bar{\theta}_{23}$ \cite{ANUMU}.  The corresponding search for $\bar{\nu}_e$ appearance is ongoing. 

\section{Proton accumulation}
The T2K experiment was designed to accumulate $7.8 \times 10^{21}$ POT.  The accumulated POT is expected to increase linearly each year until around 2018, when an upgrade to the main ring at J-PARC is expected to take place.  This planned power supply upgrade would increase the repetition rate and result in a higher beam power of around 750 kW.   Subject to this upgrade in 2018, the design goal of $7.8 \times 10^{21}$ POT should be achieved by 2020/2021.  The acquisition of the full POT of $7.8 \times 10^{21}$ has been fully approved by the J-PARC physics advisory committee. 

To June 2015, T2K has recorded around 15\% of its end goal POT.  Based on the number of events observed by the experiment to June 2015 and assuming equal portions of neutrino and anti-neutrino running, the expected yields of each event type for the full POT may be predicted.  These are given in Table \ref{stats}.  Statistics are lower for the anti-neutrino mode due to the high, intrinsic $\nu_{\mu}$ background in the beam (around 30\%).   As the disappearance probability depends on $\delta_{CP}$, two scenarios for each run configuration are given.  It should be noted that there is a small, but not statistically significant, difference between the projected $\nu_{e}$ appearance statistics for the two values of $\delta_{CP}$ quoted in Table \ref{stats}. 
\begin{table}[t]
\begin{center}
\begin{tabular}{l|ccccccc}  
 &  $\delta_{CP}$ & Total & Signal $\nu_{e}$ & Signal $\bar{\nu}_e$  & Beam $\nu_e$ & Beam $\nu_{\mu} $ & NC \\ 
\hline 
 \begin{tabular}{l}Neutrino \\mode\end{tabular} & \begin{tabular}{c}0 \\-90\end{tabular} & \begin{tabular}{c}145.8\\170.9\end{tabular} & \begin{tabular}{c}106.0 \\131.4\end{tabular} & \begin{tabular}{c}1.2 \\0.8\end{tabular} & 20.6 &0.7 & 17.2 \\
\hline
 \begin{tabular}{l}Anti neutrino \\mode\end{tabular}  & \begin{tabular}{c}0 \\-90\end{tabular} & \begin{tabular}{c}47.5\\41.5\end{tabular} & \begin{tabular}{c}5.6 \\6.5\end{tabular} & \begin{tabular}{c}24.4 \\17.5\end{tabular} & 8.6 &0.2 & 8.6\\
\hline
\end{tabular}
\caption{Projected statistics for each event type observed at Super-Kamiokande for the T2K design goal POT of $7.8 \times 10^{21}$ \cite{FST}. Note that NC refers to Neutral Current background events.}
\label{stats}
\end{center}
\end{table}

The small difference can be translated into a limit for the discovery of $CP$ violation and sensitivity to $\delta_{CP} \neq 0$.  Assuming the least sensitive case (i.e. normal hierarchy), T2K has $>90\%$ C.L. sensitivity if $\delta_{CP} = -90^{\circ}$.  Combining results from the T2K and NO$\nu$A experiments, would improve this limit and overall sensitivity to $\delta_{CP}$.  Assuming that both the T2K and NO$\nu$A experiments each take 50:50 neutrino and anti-neutrino data, sensitivity to $\delta_{CP}$ is improved further still, but not sufficiently to claim a $5 \sigma$ discovery.  However, a combined analysis is likely to indicate the most likely value of $\delta_{CP}$ and discovery by the next generation of long-baseline neutrino experiments should be possible.

\section{T2K prospects post 2020/2021}
The T2K beam will achieve its design goal POT and beam power in 2020/2021.   To this end, a high-intensity beam study was conducted in June 2015.  The study used 2 bunches per spill instead of the standard 8 bunches per spill used in T2K operations.  Results from these trials were extrapolated to determine the maximum beam power for different repetition rates of the main ring.   The high repetition rate, i.e. repetition period of 1.3 seconds, will be required to achieve a 750 kW beam.  Interestingly, this repetition rate would also be capable of producing a 1 MW beam, subject to replacing the magnet power supplies, RF cavities and miscellaneous injection and extraction equipment of the main ring.  However, as several of these upgrades would be required to achieve the 750 kW beam,  achieving a 1 MW beam should be relatively straightforward.

The components of the T2K neutrino beamline (e.g.~target, horns, decay volume, beam dump) were constructed to withstand a beam power in excess of 1 MW.  In addition there are no beam power restrictions at both the near and far detectors.  Therefore, with no limiting factor, the T2K experiment would be able to accept a 1 MW beam whenever it is available. 

\begin{figure}[htb]
\centering
\includegraphics[height=10cm]{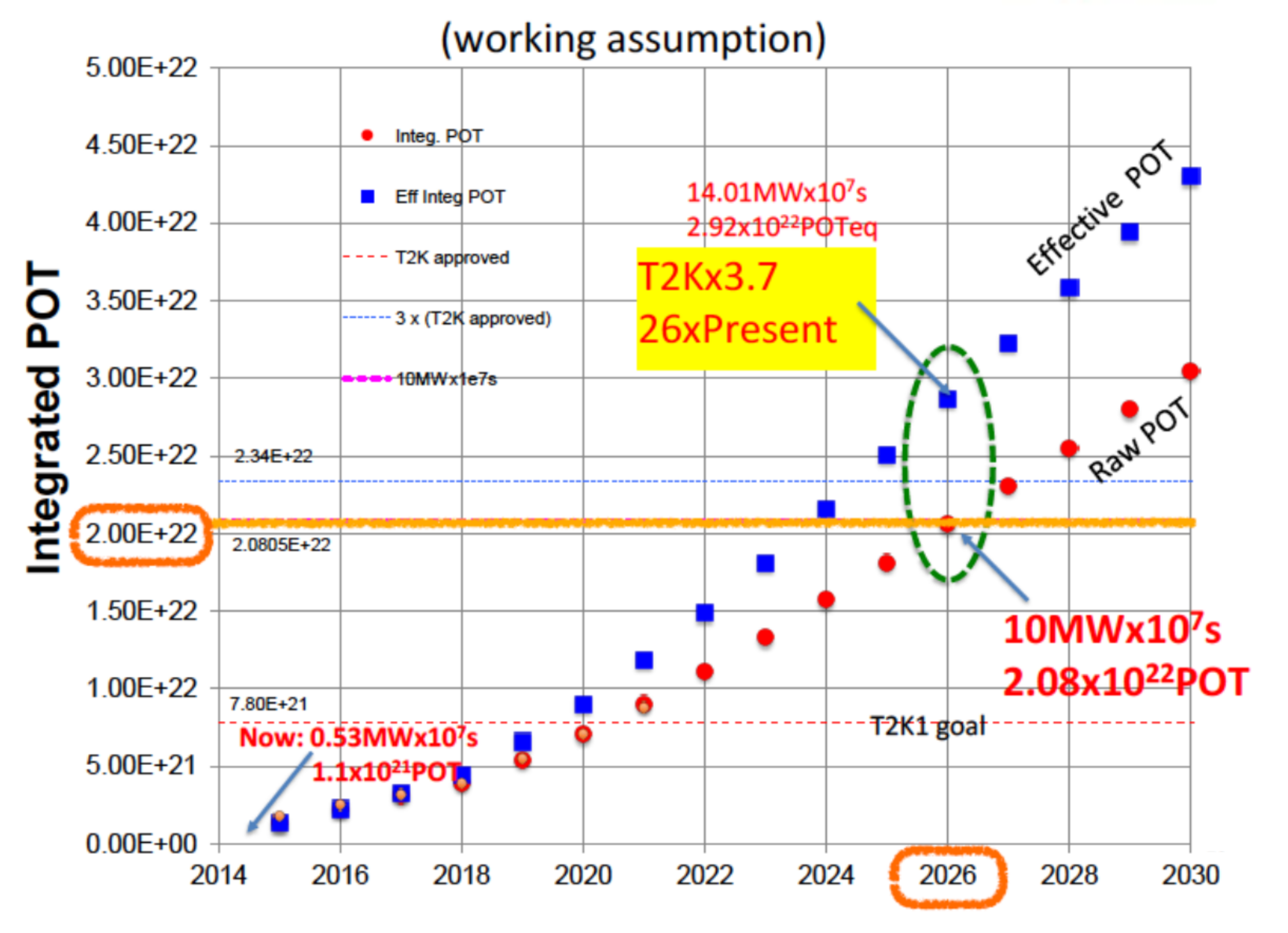}
\caption{POT projection for T2K (to 2020/2021) and beyond.  A main ring power supply upgrade in 2018 is assumed.}
\label{P2020}
\end{figure}

The possibility of running the T2K experiment with a 1 MW beam presents an interesting prospect.  The next generation of long-baseline neutrino experiments (e.g. DUNE and Hyper-Kamiokande) are not expected to commence operations until 2026, which potentially leaves a 5 year gap between their start and the end of operations at T2K in 2021. A 1 MW neutrino beam at J-PARC would provide an opportunity to bridge a gap in the worldwide long-baseline neutrino physics programme by extending operation of the T2K experiment.   The post-2020 POT projection is shown in Figure \ref{P2020} for two different scenarios.  The difference between the red and blue points reflect that improvements to the efficiency of the near detectors will be improved via improvements in the reconstruction and analysis techniques and the analysis sample at the far detector will be increased by including non-CCQE interactions.  This will give an increased effective POT relative to the current status of the experiment. 

The proposal to extend the operation of the T2K experiment beyond its design goal is begin considered  by the collaboration.  Before 2025, a $>1$ MW neutrino beam from J-PARC could provide $20 \times 10^{21}$ POT to the T2K experiment.  With improvements to detector efficiency and hardware, this would be an effective POT of $25 \times 10^{21}$.  Assuming a 50:50 split between neutrino and anti-neutrino mode, the projected statistics for the T2K design goal and extended running POTs are given in Table \ref{end}.  An effective POT of $25 \times 10^{21}$ increases the significance of the difference between the $\nu_e$ appearance for the two cases given in Table \ref{end}, leading to an improvement in the sensitivity to $\delta_{CP}$.  For an effective POT of $25 \times 10^{21}$ a $3\sigma$ determination of $CP$ violation using T2K alone would be possible given the current accuracy of $\theta_{23}$. However, given further improvement in the measurement of $\theta_{23}$,  sensitivity to $CP$ violation and $\delta_{CP} \neq 0$ would improve slightly.  The additional statistics from extended running would help improve the accuracy of the T2K $\theta_{23}$ measurement.  

For extended running with a 50:50 neutrino anti-neutrino running, $3 \sigma$ sensitivity should be possible for around 30\% of $\delta_{CP}$ values.  Systematic errors of around 2-3\% would be desirable and further improvements in the beamline and analysis methodologies would further improve the sensitivity.   For example, adjustment of the horn current would increase sensitivity for appearance and disappearance measurements as a higher horn current reduces backgrounds in the beam.    $\delta_{CP}$.  With an increased horn current of $\pm320$ kA, a $3\sigma$ sensitivity would be achieved more quickly when compared with a nominal horn current of $\pm250$ kA. In addition, if the mass hierarchy is known, the sensitivity to $CP$ violation increases further to $3\sigma$ for 45\% of the full $\delta_{CP}$ region.  

\begin{table}[t]
\begin{center}
\begin{tabular}{c|ccccc}  
POT &  $\delta_{CP}$ & Signal $\nu_{e}$ &  Beam $\nu_e$  & Signal $\bar{\nu}_e$ & Beam $\bar{\nu}_{e} $ \\ 
\hline 
 \begin{tabular}{l}$7.8 \times 10^{21}$ \\ \end{tabular}  & \begin{tabular}{c}0 \\-90\end{tabular} & \begin{tabular}{c}98.2\\121.4\end{tabular} & \begin{tabular}{c}26.8 \\26.4\end{tabular} & \begin{tabular}{c}25.6 \\19.0\end{tabular} & \begin{tabular}{c}16.3 \\17.2\end{tabular}  \\
\hline
 \begin{tabular}{l}$25 \times 10^{21}$ \\(effective)\end{tabular}  & \begin{tabular}{c}0 \\-90\end{tabular} & \begin{tabular}{c}314\\389\end{tabular} & \begin{tabular}{c}85.9 \\84.6\end{tabular} & \begin{tabular}{c}82.1 \\60.9\end{tabular} &  \begin{tabular}{c}52.2 \\55.1\end{tabular}\\
\hline
\end{tabular}
\caption{Projected statistics for T2K design goal POT and projected POT for extended T2K running to 2025 with a 1 MW beam (post 2020/2021).}
\label{end}
\end{center}
\end{table}

\section{Conclusions}
The T2K experiment made the first observation of $\nu_e$ appearance in a $\nu_{\mu}$ beam at a baseline of 295\,km and peak beam energy of 0.6\,GeV.  By combining the T2K result with the world average value of $\theta_{13}$ from reactor experiments, $\delta_{CP}$ between $0.19\pi$ and $0.80\pi$ was ruled out at 90\% C.L.~for the normal hierarchy. For the inverted hierarchy values between $-\pi$ and $-0.97\pi$ and $-0.04\pi$ and $\pi$ are excluded at 90\% C.L.   The T2K experiment will continue to take data and investigate $CP$ violation in the lepton sector more precisely, but even combination with results from other experiments will not yield a discovery, even for the most favourable $\delta_{CP}$ value.

The design goal of the T2K experiment is to collect the approved POT ($7.8 \times 10^{21}$) data by around 2020, to conduct precision studies of neutrino oscillations with neutrino and anti-neutrino beams.  Around 2020, there is a possibility that the J-PARC facility could produce a 1 MW or above neutrino beam. An extension of T2K (currently called T2K-II) is under discussion and consideration by the T2K collaboration. 

Between 2021 and 2025, T2K could collect $20 \times 10^{21}$ POT with 1.3 MW beam power, with which we will have $3\sigma$ CP violation discovery sensitivity for $\delta_{CP} = -\pi/2$. By improving the T2K efficiency, it may be possible to enhance the effective POT to $25 \times 10^{21}$.  A 1 MW beam neutrino beam at J-PARC would provide exciting opportunities for extending the running of the T2K experiment (T2K-II) and would have have a positive impact for the next generation experiment Hyper-Kamiokande, as it ensures the presence of a high powered beam from the first day of operation.

\end{document}

%% file: econfmacros.tex
\def\beq{\begin{equation}}
\def\eeq#1{\label{#1}\end{equation}}
\def\eeqn{\end{equation}}

\def\beqa{\begin{eqnarray}}
\def\eeqa#1{\label{#1}\end{eqnarray}}
\def\eeqan{\end{eqnarray}}

\let\bar=\overbar

\def\Dslash{\not{\hbox{\kern-4pt $D$}}}
\def\dslash{\not{\hbox{\kern-2pt $\del$}}}

\def\msb{{\bar{\ssstyle M \kern -1pt S}}}